\RequirePackage{fix-cm}
\documentclass[twocolumn]{svjour3} 

\usepackage{caption}
\usepackage{csquotes}
\usepackage{amsmath,amssymb,amsfonts}
\usepackage[sorting=nyt,style=apa,backend=biber]{biblatex}
\usepackage[symbol]{footmisc}

\usepackage{subcaption}
\usepackage[section]{placeins}
\usepackage{balance}
\usepackage{xcolor}
\usepackage{makecell}
\usepackage[normalem]{ulem}
\usepackage{orcidlink}
\RequirePackage{graphicx}
\journalname{}

\begin{document}

\title{Network Modelling in Analysing Cyber-related Graphs}
\titlerunning{Network Modelling in Analysing Cyber-related Graphs}        

\author{Vesa Kuikka\orcidlink{0000-0002-3677-816X}\thanks{Corresponding author}\thanks{e-mail: vesa.kuikka@aalto.fi}
        \and
        Lauri Pykälä\thanks{}
                \and
        Tuomas Takko\thanks{}
                \and 
        Kimmo K. Kaski\orcidlink{0000-0002-3805-9687}\thanks{}
}

\institute{Department of Computer Science, Aalto University School of Science,
P.O. Box 11000, 00076 Aalto, Finland \label{addr1}
}

\date{Received: date / Accepted: date}

\maketitle

\begin{abstract}

To improve the resilience of the computer network infrastructure against cyber attacks or causal influences and find ways to mitigate their impact, we need to understand their structure and dynamics. Here, we propose a novel network-based influence-spreading modelling approach to investigate event trajectories or paths in attack and causal graphs with directed, weighted, cyclic and/or acyclic paths. In our model, we can perform probabilistic analyses that extend beyond traditional methods to visualise cyber-related graphs. The model uses a probabilistic method to combine paths that join within the graph. This analysis includes vulnerabilities, services, and exploitabilities. To demonstrate the applicability of our model, we present three cyber-related use cases: two attack graphs and one causal graph. This model can serve cyber analysts as a tool to produce quantitative metrics for prioritising tasks, summarising statistics, or analysing large-scale graphs.


\end{abstract}

\section{Introduction}\label{sec:Introduction}

Cyber attacks are unauthorised actions against computer network infrastructure to exploit vulnerabilities in target systems, while cyber threats refer to potential adversarial events in which an attacker could exploit vulnerabilities (\cite{li2021comprehensive}). Typically, a cyber attack can consist of a number of actions, events, or steps required for the attacker to gain access or sufficient privileges to achieve the objective of an attack, e.g. denial of service (DOS) or the delivery and use of malware. A common approach to classifying these steps has been through various frameworks such as the cyber kill chain (\cite{hutchins2011intelligence}).
At a more technical level, potential or real cyber attacks have often been modelled as directed graph structures. Among these graph-based methods, attack graphs and their variants are often
used to represent cyber attacks (\cite {lallie2020review,zenitani2023attack,wachterGraphModelsCybersecurity2023,zeng2019survey,liu2012using}). There are two commonly used forms of attack graphs: the first is a directed graph with nodes representing network states and edges (links) representing exploits, while the second uses nodes to represent pre- or post-conditions of an exploit and edges to show the consequences. In addition, knowledge-based methods are used to model attack concepts and vulnerability details (\cite{alserhani2015knowledge,qi2023cybersecurity,sikos2023cybersecurity}).

In order to understand, analyse, and visualise the sequence of events of a successful cyber attack, we will here use attack modelling techniques (AMTs). There are three main categories of AMT: (i) methods based on use cases like misuse or security use cases, (ii) temporal methods based on e.g. cyber kill chains, and (iii) methods based on graphs (\cite{lallie2020review}). Here, we will focus on graph-based models as they enable cyber analysts and researchers to formulate and use different metrics, both quantitative and qualitative, to optimise and prioritise defensive efforts (\cite{lallie2020review}).
Graph-based cyber attack modelling can be conducted at various technological levels, including applications, services, or entire computer networks. To achieve this, mappings from technical micro-service levels to higher clustering levels can be accomplished using technical configuration repositories.

In the literature, the term \emph{attack graph} has been used to refer to a variety of modelling approaches (\cite{wachterGraphModelsCybersecurity2023}), which present a taxonomy of these models based on analysing $70$ attack graph formalisms. This taxonomy uses a hierarchical categorisation 
with two top-level categories: sequential models and causal dependency models. Sequential models represent attack steps and paths formed by them, while causal dependency models focus on cause-effect relationships. Based on node semantics, sequential models are further categorised into asset-oriented, vulnerability-/exploit-oriented, and condition-/attribute-oriented types.

Our research approach will focus on network modelling, as it provides a range of methods and tools for analysing cyber-related graphs from different perspectives and aggregation levels. One recent approach involves modelling network structure and influence flow at a detailed level of nodes and links. (\cite{kuikkaSciRep,Almiala_Aalto_Kuikka,Kuikka_Aalto}) This network model uses Markov processes and applies probability theory to assess the combined effects from multiple sources. Our method for evaluating the effects of various joining paths in both acyclic and cyclic (\cite{levner2024fast}) graphs extends the probabilistic techniques discussed in the literature. (\cite{wang2008attack,stergiopoulos2022automatic,carter2014probabilistic,homer2013aggregating}) 
In this study, we calculate the combined effect of multiple alternative paths in a graph by using the theorem of non-mutually exclusive events from probability theory. Details of the model and its corresponding algorithm can be found in (\cite{Kuikka_Aalto}).
In general, our probabilistic modelling approach serves as an integrated, graph-based methodology or framework for analysing various types of influence spreading processes, including cyber attacks. The integration of modelling and analysis metrics within the same probabilistic framework makes our approach self-consistent, distinguishing it from other graph-based methods.

In this study, we present the modelling of cyber-related graphs through three use cases and demonstrate how the model can be applied to examine the exploitability, causality, and incomplete graph information in the computer network and services system. The first use case (\cite{stergiopoulos2022automatic}) involves a directed acyclic graph in which the states of the system are represented as nodes and the vulnerabilities as links between the nodes, forming an attack graph. The exploitability of a vulnerability is interpreted as the probability of a successful exploit (i.e., traversability of a link). In the network model, these probabilities 
are assigned to link weights, which are derived using general CVSS scoring data (Common Vulnerability Scoring System)~(\cite{joint2012nist}) and adjusted with a scaling factor. The second use case (\cite{alsaheel2021atlas}) demonstrates a larger network structure of services and applications, illustrating how exploitability metrics can be utilised to produce aggregate security metrics concerning vulnerability and service levels. The third use case (\cite{alsaheel2021atlas,genweijiangCVE20170199WildAttacks}) involves two types of nodes, namely attack entities and non-attack entities identified by an attack investigation tool using natural language processing and deep learning. This use case illustrates how our model can address inaccuracies in results from threat analysis tools.

These use cases demonstrate how different categorisations can be applied within the model. Because the model presents metrics as probabilities, it aids in forming a consistent understanding of the situation from various perspectives. Additionally, our model introduces novel methods for analysing cyber-related graphs. It employs a probabilistic technique to merge multiple alternative paths—both cyclic and acyclic—within a graph. The model uses a detailed model to consider all potential paths through the nodes of the network and defines new metrics to estimate the exploitability and impacts. For instance, it can be used to evaluate the impact of eliminating certain vulnerabilities across services or protecting targeted services within the system. 

Therefore, our model could serve as a valuable aid for cyber analysts and those designing cyber-secure systems. Based on our model, it is possible to develop a visualisation and analysis tool that allows cyber analysts to leverage various model outputs and metrics for real-world mitigation planning. This tool could include enterprise-specific assessments related to exploitability and impact values. As a result, analysts could perform what-if analyses and inspect specific parts of networks and systems.

In Section~\ref{sec:Model}, we will present our network modelling method and analysis metrics.
The pseudo-algorithms of our influence spreading model are reproduced in the supplementary material from our earlier study~(\cite{Kuikka_Aalto}). In Section~\ref{sec:RelatedWork}, we review related research on attack graphs and metrics used to assess the impacts of cyber attacks. In Section~\ref{sec:Data}, we describe the data and networks of the use cases on which the model is demonstrated. In Section~\ref{sec:Results}, we will explain how the model can be used to obtain new results in analysing cyber-related graphs through the three use cases. In Section~\ref{sec:Conclusions}, we will discuss the applications of our graph-based methods in analysing cyber vulnerabilities and attacks, and present concluding remarks.
\section{Related Work}
\label{sec:RelatedWork}

Several reviews have been published about attack graphs and related cybersecurity methods and models to analyse them (\cite{lallie2020review,wachterGraphModelsCybersecurity2023,zenitani2023attack,zeng2019survey}). The review in (\cite{lallie2020review}) provides a description of a theory of cyber attacks and how elements of attack graphs and attack trees are represented and visualised. The study in (\cite{wachterGraphModelsCybersecurity2023}) presents the state of research on the representation and analysis of cyber attacks 
using attack graph formalisms and proposing a classification based on an analysis of models regarding graph semantics, agents involved,
and analytical features. This study also investigates which formalisms enable automatic attack graph generation from raw or processed data input.

The review in (\cite{zenitani2023attack}) presents a recent summary of studies on attack graphs, by focusing on key concepts such as exploit dependency and / or rules (\cite{levner2024fast}), monotonicity, and handling cycles. The study in (\cite{zeng2019survey}) describes the key concepts, generation methods, and computation tasks of attack graphs. Also, it summarises various analysis methods, including graph-based, Bayesian network-based, Markov model-based, cost-optimisation, and uncertainty analysis methods. In the attack graph analysis, there is a need to quantify the levels of threats,  vulnerabilities, exploits, and impact, each with appropriate metrics. For example, in enterprise environments, system security metrics (\cite{pendleton2016survey}) and network attack graph metrics (\cite{Noel2017}) have been proposed to evaluate and compare the threats and impacts of cyber attacks.

In the literature (\cite{Shakarian2015}), an Independent Cascade (IC) model has been proposed as a way to describe influence maximisation. This model assumes that nodes activate their neighbours independently on the basis of a fixed probability rather than multi-step logical relationships, as in the case of our influence spreading model. A key difference is that the IC model does not consider joining paths, like our present model does by following the probabilistic rule of mutually non-exclusive events. This can result in a more realistic evaluation of the overall risk for the enterprise network and services. Our probabilistic method also allows us to calculate expected cumulative and non-cumulative impact values, as well as the circular effects of attack propagation.

The present study is related to two earlier works: one by Stergiopoulos et al. (\cite{stergiopoulos2022automatic}) and another by Alsaheel et al. (\cite{alsaheel2021atlas}). The data for the first two use cases are adapted from (\cite{stergiopoulos2022automatic}), while the data for the third use case comes from (\cite{alsaheel2021atlas}). The study in (\cite{alsaheel2021atlas}) presents a sequence-based learning approach to investigate attacks, whereas  (\cite{stergiopoulos2022automatic}) proposes an automatic analysis of attack graphs to facilitate risk mitigation and prioritisation using probabilistic methods.

Our method uses probability-based path combination, which complements the approach outlined in (\cite{stergiopoulos2022automatic}). In that approach, single paths, as well as those with minimum and maximum exploitability and impact values, are assessed. However, our method is well
suited for the comprehensive evaluation of enterprise networks. In contrast, the single path method allows for a detailed analysis of individual attack paths. Both techniques can be used together to examine different aspects of cyber attacks. Moreover, in enterprise-scale networks, analysing single paths can be computationally resource-intensive due to the vast number of potential combinations. Relying solely on subjective judgment to limit the set of possible paths may not produce accurate results.

 Although there are a number of studies using graph-based cyber attacks modelling, a comprehensive approach applicable to various levels of applications and services in enterprise networks is still lacking. In the present study, we introduce a probabilistic method to quantitatively analyse cyber-related graphs and identify strategies to improve the resilience of network infrastructure.
\section{Attack Graph Model}
\label{sec:Model}
Graphs related to cyber security, such as attack graphs and causal graphs, are mainly created to help visualise complexities of attack sequences and plan ways to prevent them. (\cite{lallie2020review}) These graphs are usually generated from large volumes of data for this purpose (\cite{sheyner2003tools,stojanovic2020apt}). Consequently, the resulting graphs typically contain a small number of nodes, ranging from a few to hundreds.

In our network-based modelling approach (\cite{Kuikka_Aalto}) we consider and analyse cyber attack processes from the point of view of influence spreading for finding ways to prevent or mitigate potential or ongoing  attacks. (\cite{haque2021realizing,segovia2024survey}) The model describes the network structure, including individual nodes and links. Influence spreads through a non-conserved process, moving from a node to all neighbouring nodes via directed links. In our model, the influence spreading can be assumed to take place via acyclic or cyclic Markov processes on a complex network structure.  
In the present study, the model describes attack propagation along self-avoiding paths, representing acyclic paths, and uses unrestricted paths to represent general propagation through either acyclic or cyclic paths.

The weights assigned to directed links represent the probabilities of successful attack propagation steps within a graph structure. Since the actual parameter values, including link weights and CVSS scores, can vary greatly in real environments, in this study, we employ a scaling factor to cover the full range of possible parameter values. This approach also enables what-if and sensitivity analyses to explore various attack scenarios and environments.

Our model is quite versatile as it offers different approaches to dealing with full breakthrough of nodes or self-avoiding paths. The first approach allows loops in the spreading process, while the second approach is better for examining connectivity between nodes. Loops consisting of a minimum of three nodes connected by links enable circular propagation of influence within the network structure. Both approaches yield similar results for directed acyclic graphs because the network lacks cyclic processes. In case of full breakthrough effects, a bidirectional link creates a cycle between the two nodes at each end of the link, resulting in recurring events in the network. Note that due to spreading, the effects are not limited to those nodes but spread throughout the network structure. In the case of self-avoiding paths, no recurring effects are permitted, and the process can propagate only in one direction along a path.

The method and corresponding algorithms proposed in our earlier article (\cite{Kuikka_Aalto}) are based on modelling network structure and spreading processes. This results in an $N \times N$ probability matrix $C$, where the number of nodes in the network is denoted by $N$. The matrix elements represent directed spreading or connectivity probabilities from one node to another in the network with a connecting path between them. When spreading from branching paths, it occurs in all possible directions, and the probabilities concern the first arrival at the end nodes. Subsequently, this matrix is used to define various metrics for analysing the network structure and network flow process. Note that
models other than our influence spreading model can generate the probability matrix. As the analysis is based on the probability matrix, it is independent of the method used to produce the probability matrix. These models could even be non-Markovian or include other event types besides mutually non-exclusive events (OR rule) used to combine multiple joining alternative paths (\cite{Kuikka_Aalto}).

In network analysis, we use two metrics to evaluate the importance of a node, namely out-centrality and in-centrality, and they are defined differently from the standard closeness centrality measures in the literature. One reason for using different metrics is the limitations of standard metrics in describing detailed network structures, because they are defined based on the shortest paths between nodes, thus ignoring other alternative paths an attack could propagate. (\cite{critical}) We use the probability matrix as the basis of analysis, as it enables the separation of the network modelling from the analysis. (\cite{Kuikka_Aalto})

Next, we define the centrality measures for the subset of nodes $V$ in the network $G$ of $N$ nodes. (Note that in the case of the entire network $V$=$G$). For the influence spreading through the network, we define the out-centrality of  source node $s$ by 
\begin{equation}
\label{eq:out}
C^{(\mathrm{out})}(s)=\frac{1}{N-1}\sum_{\substack{t\in V \\ s \neq t}}C(s,t)
\end{equation}
and the in-centrality of a target
node $t$ by
\begin{equation}
\label{eq:in}
C^{(\mathrm{in})}(t)=\frac{1}{N-1}\sum_{\substack{s\in V \\ s \neq t }}C(s,t).
\end{equation}
Here, the out-centrality of a node is the average value of probabilities of spreading from the source node to all other nodes in $V$, while the in-centrality of a node is the average value of probabilities of spreading from all other nodes in $V$ to the target node. These centrality measures describe the physical properties of the network structure and the flow process. As $C(s, t)$ is the probability of influence spreading from node $s$ to node $t$, the sum in Equation~\ref{eq:out} is interpreted as the expected value of the influenced nodes and the sum in Equation~\ref{eq:in} as the expected number of nodes that spread influence to the specified node. In the context of cyber-related graphs, influence spreading refers to the propagation of cyber attacks within the graph. In this study, we adopt the convention of setting the state value of the source node as zero and using the corresponding normalisation factor $1/(N-1)$ and express the results as percentage values.

When dealing with cyber-related graphs, the out-centrality is a measure for assessing the potential effect of a cyber attack on other nodes within the network. On the other hand, in-centrality is useful for understanding how various cyber attacks can affect nodes in the network. Detailed information can be obtained by focusing on a subset of start and end nodes using Eqs.~\ref{eq:out} and \ref{eq:in}. Typically, the main focus is to explore vulnerabilities and the effect of a start node on an end node. However, multiple start nodes and end nodes can exist in one cyber-related graph.

Let us move on to discuss the metrics of exploitability and impact. The exploitability represents the probability of a successful cyber attack through a link, while impact describes its effect on the system, functionality, or operation. To assess the impact of different needs both at the technical and operational levels, the first step is to define the concept in the specific situation. The impact is calculated for the nodes and the exploitability is calculated for the links. We assume that the impact value does not influence the probabilities of propagation of the cyber attack. Therefore, impact values may not be expressed as probabilities and can have numerical values beyond the range of $[0, 1]$. However, the impact values for different nodes are comparable, enabling us to calculate the desired sums for paths or network structures.

When assessing vulnerabilities, there are two main approaches: the General CVSS scoring (Common Vulnerability Scoring System)~(\cite{joint2012nist}) or a custom evaluation method in which organisations supplement or override CVSS with their own risk assessments. The empirical CVSS scoring system provides a quantitative way to capture key characteristics of a vulnerability, resulting in scores that indicate its severity (the impact on a system) and exploitability (the ease of exploitation).~(\cite{Scoring}) These metrics adhere to international standards for measuring cybersecurity risks. By using either CVSS or a custom evaluation, organisations can prioritise patches and mitigation strategies based on the potential impact of each vulnerability. We demonstrate the use of the CVSS scoring system in our first use case. In our second use case, since the general CVSS scores are uniformly high across all vulnerabilities, we focus on illustrating the network structure of services and applications and, therefore, use equal values for the vulnerability characteristics of all vulnerabilities. Furthermore, it is important to note that the general CVSS method may not accurately reflect the actual risk in a specific organisation or environment. The equations and algorithms used to score the base, temporal, and environmental metric groups are detailed in~(\cite{Scoring}). Additional information on the origin and testing of these equations can be found at~(\cite{FIRST}).

Our approach is based on the use of link weights as probabilities to represent successful exploits. The exploitability metric measures the current state of exploit techniques or code availability. When easily accessible exploit code is publicly available, it increases the number of potential attackers, including unskilled ones, increasing the severity of vulnerabilities. The following equation illustrates how the factors $AccessVector$,  $AccessComplexity$, and $Authentication$ describe the accessibility and complexity of the vulnerability, and whether additional conditions are needed to exploit it. The exploitability metric is defined as follows (\cite{Scoring}):
\begin{multline}
\begin{aligned}
Exploitability = 20*Access & Vector* \\  AccessComplexity & *Authentication.
\label{eq:ScoExp}
\end{aligned}
\end{multline}

Impact metric measures the potential consequences of exploiting a vulnerability of an IT asset. These impacts are independently defined based on the degree of loss in three key areas, namely confidentiality, integrity, and availability. The present analysis can be further extended to consider specific business impacts by incorporating relevant effects in Equation~\ref{eq:impact}. The total impact metric is defined as follows (\cite{Scoring}):
\begin{multline}
\begin{aligned}
Impact = 10.41 & *(1-(1-ConfImpact)* \\(1-Integ & Impact)* 
(1-AvailImpact)).
\label{eq:ScoImp}
\end{aligned}
\end{multline}
In a network structure, a node can be targeted by different cyber attack paths from the source node through various links directed to the target node. In an attack graph, these may involve the same vulnerability. However, if different vulnerabilities can be exploited on one node, we calculate the weighted average value for the node or use the maximum impact of alternative exploits. We define a vector $I$ with elements representing the impact values for the nodes in the network. The impact of a cyber attack from the start nodes $S$ to the end nodes $T$ can be defined as:

\begin{equation}
    \label{eq:impact}
    \mathcal{I}(S,T;I)=\sum_{\substack{i \in S \\ j \in T}}C(i,j)I_j,
\end{equation}
where the matrix $C$ consists of elements $C(i,j)$ that represent the probabilities of successful attacks from node $i$ to node $j$ via alternative paths. The impact value on the end node alone includes the last exploit on the path. The impact value of Equation~\ref{eq:impact} can be calculated for specific events, including effects on specific services or aspects of security. Additionally, Equation~\ref{eq:impact} can be utilised to analyse the impact of exploits by determining the decrease in the impact value after mitigating vulnerabilities.

When all elements of vector $I$ are set to one, we obtain a quantity that describes the attack propagation from the source node set $S$ to the target node set $T$ in the network. In this way, the impact value in Equation~\ref{eq:impact} is linked to the out- and in-centrality metrics in Eqs.~\ref{eq:out} and \ref{eq:in}. Therefore, the centrality measures and the impact measure in Equation~\ref{eq:impact} are defined consistently with each other. Combining the alternative paths using probabilistic methods allows us to calculate accurate metrics based on Eqs.~\ref{eq:ScoExp} and \ref{eq:ScoImp} for real-world cyber attacks and different classifications.

As a further characteristic measure, one can calculate a cumulative impact for an individual attack path, as done in (\cite{stergiopoulos2022automatic}). In our model approach, the cumulative impact along a path is calculated by summing up the products of propagation probabilities and node impacts according to Equation~\ref{eq:impact}. Once again, the impact on the end node only includes the impact on that node.

In summary, four kinds of impact metrics can be calculated: cumulative impacts along individual paths or all alternative paths, and non-cumulative impacts on the end nodes through individual or all alternative paths. In this study, we prefer metrics that combine alternative paths using a probability theory approach and methods. The factors $C_{i,j}$ in Equation~\ref{eq:impact} account for the effects of all potential alternative paths. This approach allows for the definition of new metrics in analysing cyber events from different perspectives and categorisations. In the next section, we provide examples of this through use cases.
\section{Use Case Data}
\label{sec:Data}

For this study, we chose three different graph structures, described in Table~\ref{tab:graphs_summary} to serve as our use cases.

\begin{table*}[ht]
    \centering
    \caption{Summary of the use case graphs.}
    \begin{tabular}{|c|c|c|c|c|} \hline 
 Graph& Description& Nodes& Links
&Reference\\ \hline \hline 
          Multi-cloud Enterprise Network&  Netflix OSS microservice system&  18&  
25& \cite{stergiopoulos2022automatic}\\ \hline 
         Netflix OSS&  Netflix OSS microservice system&  21&  94& \cite{stergiopoulos2022automatic} \\ \hline 
         ATLAS, Pony campaign attack&  Recovered sequences and a causal graph&  
39&  65& \makecell{\cite{alsaheel2021atlas} \\ \cite{genweijiangCVE20170199WildAttacks}} \\ \hline
    \end{tabular}
    \label{tab:graphs_summary}
\end{table*}

\paragraph{\textbf{Use case 1: Multi-cloud Enterprise Network.}}
The graph for the Multi-cloud Enterprise Network from  (\cite{stergiopoulos2022automatic}) represents the network topology of two cloud infrastructures connected to the Internet behind an external firewall. The first cloud server hosts three virtual machines, all connected to a virtual switch. The second cloud server hosts a public network and a private network. External users can access a web server, and internal users can access the SQL server from inside the local network.
Each server in this topology represents a realistic vendor-specific system with a set of real-world CVE vulnerabilities, with the impacts of exploitation extracted from the CVE database.

\paragraph{\textbf{Use case 2: Netflix OSS architecture.}}
\cite{ibrahimAttackGraphGeneration2019} used a combination of tools to construct attack graphs for microservice architectures, consisting of a Docker host running various interconnected containers. The network topology was extracted from the docker-compose configuration files that define the orchestration of the services and details about the connections between containers, published ports, and any privileged access granted to containers.
Vulnerabilities were scanned from the Docker images using the vulnerability scanning software Clair, producing a list of CVEs with textual descriptions and attack vectors for each image. The topology and vulnerability data were used as input for the final attack graph generation process. Attack pre- and postcondition parsing was performed by matching the attack vectors of the vulnerabilities by keyword matching \cite{aksuAutomatedGenerationAttack2018}.

\paragraph{\textbf{Use case 3: Pony APT (Advanced Persistent Threat)} campaign.}
In the research conducted on the Pony APT campaign (CVE-2017-0199)  (\cite{genweijiangCVE20170199WildAttacks}), the methodology was based on the work by  (\cite{alsaheel2021atlas}), describing a framework for constructing "attack stories" from audit logs. Their approach involved several steps. First, a platform independent causal graph representation was constructed using various system audit logs, including DNS records, web objects, processes, file accesses, and network connections. This graph was then abstracted using optimisation techniques to reduce its complexity. Specifically, nodes and edges that were not reachable from the attack nodes or attack symptom nodes were removed, repeated edges between entities were reduced to only the first occurrence, and nodes and edges referring to the same type of event were combined.

Next, the process of attack sequence construction took place. All \enquote{attack entities} from the causal graph were obtained, and subsets consisting of two or more attack entities were formed. The neighbourhood graph for each entity within these subsets was extracted, followed by retrieving time-stamped events for each neighbourhood graph. Sequences were labelled attack sequences if they consisted exclusively of attack events, which means that both the source and destination nodes were associated with attacks.
Following this, sequence lemmatisation was performed. This involved transforming sequences into a textual representation using a general vocabulary of 30 words, divided into four types: process, file, network, and actions. To balance the dataset, non-attack sequences were undersampled, and attack sequences were oversampled by mutating existing attack sequences. Finally, the sequences were embedded and modelled, followed by an investigation of attack and recovery of \enquote{attack stories}. In the resulting processed graph, the semantic interpretations of nodes and edges are connected to system states inferred from the audit logs. The nodes represent processes (i.e. executables), files or IP addresses. The edges between them in turn represent actions or events between the entities, like \texttt{read}, \texttt{write}, \texttt{execute}, or \texttt{connect}.

The ATLAS approach thus focuses on reconstructing the attack story from observed system behaviours, with nodes and edges closely tied to actual system events and entities as recorded in the audit logs. Currently, the calculations treat all nodes in the graph equivalently, even in the use case where the graph nodes represent different entity types. A potential direction for further study is to treat different entities in the multiplex graph as distinct and perform the calculations separately for each node class.

\section{Results}\label{sec:Results}

In this section, we present the results of our attack graph model for three different use cases. The first use case presents a small directed acyclic attack graph. (\cite{stergiopoulos2022automatic}) The second use case is a more complex attack graph from (\cite{stergiopoulos2022automatic}) and finally, the third use case is a causal graph that was generated by an attack investigation tool using natural language processing and deep learning (\cite{alsaheel2021atlas}).

\paragraph{\textbf{Use Case 1: Multi-cloud Enterprise Network.}}

We use a multi-cloud enterprise network as the first use case to demonstrate our modelling method. The network topology  
and vulnerabilities are adopted from (\cite{stergiopoulos2022automatic}). The first cloud server hosts three virtual machines: a Mail server, a Web server, and a DNS server, and the second cloud server consists of public and private networks. The public network hosts an SQL server and a NAT gateway server, and the private network hosts an Admin server and three virtual machines (VMs). Users outside the network can access the Web server, and employees within the same LAN can access the SQL server through their workstations.

\begin{figure}[ht]
     \centering
    \includegraphics[width=0.5\textwidth]{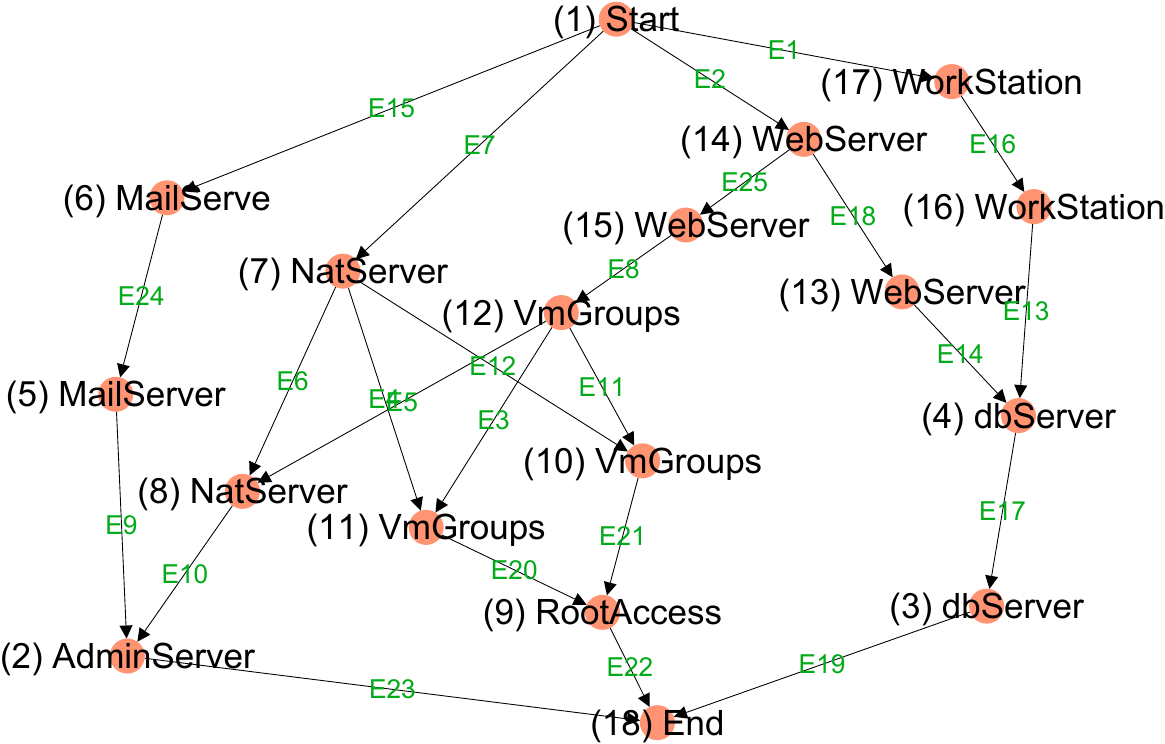}
    \caption{Use Case 1: Attack graph Multi-cloud Enterprise Network (\cite{stergiopoulos2022automatic}).}
    \label{fig:MCE}
\end{figure}

Figure~\ref{fig:MCE} shows the attack graph on the Multi-cloud Enterprise Network that consists of directed links representing the exploits and nodes representing the states. These kinds of networks are directed acyclic graphs (DAGs), i.e. without  
cycles (\cite{kordy2014dag}).

\begin{figure}[ht]
     \centering
    \includegraphics[width=0.45\textwidth]{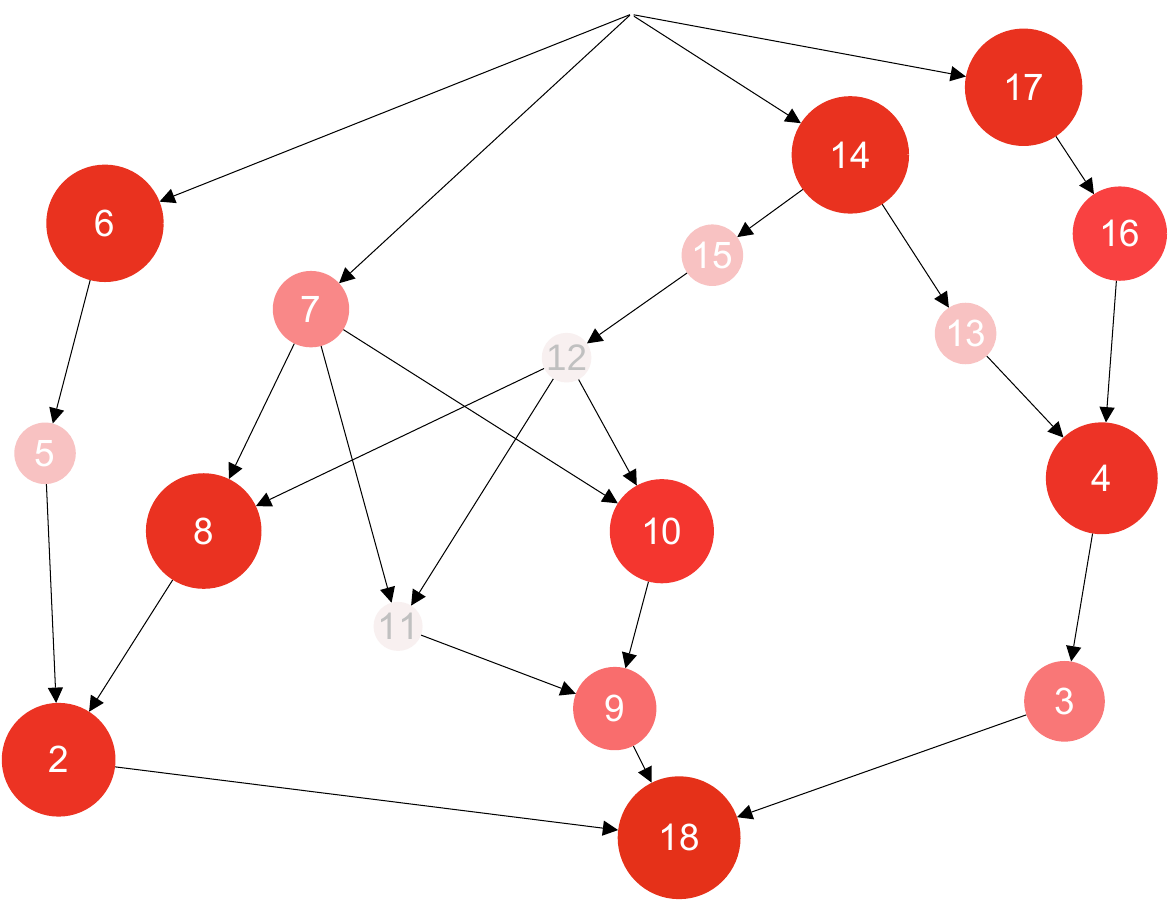}
    \caption{Impacts of successful cyber attacks from the start node $1$ of Figure~\ref{fig:MCE} to other nodes in the graph. The impact values calculated from Equation~\ref{eq:impact} are highlighted with node colours and sizes.}
    \label{fig:Impact1}
\end{figure}
The systems have vulnerabilities that are listed in the CVE vulnerabilities database. (\cite{Scoring}) The attacker's goal is to compromise a virtual machine in the private network or the database in the public network by gaining root access. The attacker can take different paths in the attack graph.

\begin{table}
\centering
\caption{Exploitability and impact values calculated from Eqs.~\ref{eq:ScoExp} and \ref{eq:ScoImp} for vulnerabilities in the attack graph of Figure~\ref{fig:MCE} (\cite{stergiopoulos2022automatic}).}
\label{tab:MCE}
\begin{tabular}{|c|c|l|p{20mm}|} \hline 
Vulnerability & Expl./$10$  &Impact 
&Links\\ \hline \hline
 -&1  &-
&E1,E2
\\\hline 
CVE-2010-3847 & 0.339258   &10.00085 
&E3,E4,E20
\\ \hline 
CVE-2003-0693 & 0.99968   &10.00085 
&E5,E6
\\ \hline 
CVE-2007-4752 & 0.99968   &6.442977 
&E7
\\ \hline 
CVE-2001-0439 & 0.99968   &6.442977 
&E8
\\ \hline 
CVE-2008-4050 & 0.85888   &10.00085 
&E9,E10
\\ \hline 
CVE-2008-0015 & 0.85888   &10.00085 
&E11,E12,E21, \newline E22,E23
\\ \hline 
CVE-2009-1918 & 0.99968   &10.00085 
&E13,E16
\\ \hline 
CVE-2018-7841 & 0.99968   &6.442977 
&E14,E18
\\ \hline 
CVE-2004-0840 & 0.99968   &10.00085 
&E15
\\ \hline 
CVE-2008-5416 & 0.7952   &10.00085 
&E17,E19
\\ \hline 
CVE-2001-1030 & 0.99968   &6.442977 
&E24
\\ \hline 
CVE-2009-1535 & 0.99968   &6.442977 &E25\\ \hline
\end{tabular}
\end{table}

In Table~\ref{tab:MCE}, we show the impact and exploitability values calculated for vulnerabilities in the attack graph (Figure~\ref{fig:MCE}). The third column of this table shows the list of links in the attack graph for each vulnerability. In this context, impacts are associated with the properties of vulnerabilities (or edges), while in Equation \ref{eq:impact} and Figure \ref{fig:MCE}, successful cyber attacks affect services (or nodes). Our model treats nodes as representations of network states; therefore, all exploits of vulnerabilities that occur on the same node are assumed to have the same impact. If this does not hold, the construction of the attack graph is not consistent, and additional network states (or nodes) should be added to the attack graph.

In Figure~\ref{fig:Impact1}, we illustrate the non-cumulative impact values of successful cyber attacks from the start node $1$ to other nodes in the graph. For example, the node $7$ has a lower impact than nodes $6, 14,$ and $17$ because the vulnerability CVE-2007-4752 from the start node to node $7$ has a relatively low impact value. On the other hand, node $10$ has a high impact value because the vulnerability CVE-2008-0015, which affects nodes $7$ and $12$ and is propagated to node $10$.
As a numerical example, the value of the expected non-cumulative impact of node $10$ is $(0.859+0.859-0.859 \times 0.859) \times 10.0 \approx 0.98 \times 10 =9.8$. The corresponding value of the expected cumulative impact is $3 \times 6.44 + 9.8 = 29.1$. This example illustrates how the probabilistic model is applied to calculate expected impacts, and the distinction between non-cumulative and cumulative impacts. The effects on other end nodes can be explained similarly according to Equation~\ref{eq:impact}. In this numerical example and Figure~\ref{fig:Impact1}, we have used a scaling factor of $w=1$ to emphasise the main idea. If a scaling factor is also used, then all exploitability values in Table~\ref{tab:MCE} are multiplied by the scaling factor to get the link weights in the attack graph model.

In our model, we use out-centrality and in-centrality metrics to analyse the propagation of attacks from potential start nodes to potential end nodes. In Figure~\ref{fig:MCE}, node $1$ is the start node and node $18$ is the end node, but there could be other nodes where the attack chain can start or end. For example, let us consider database servers (dbServer) as the attacker's goal, like in (\cite {stergiopoulos2022automatic}). Eqs.~\ref{eq:out} and \ref{eq:in} define out-centrality and in-centrality metrics as expected values over the whole network structure, not just for possible start and end nodes. There are a few reasons for this. Initially, the start and end nodes might change depending on the situation, or this information may not be available. Furthermore, complete metric data provide details about the structure of the attack graph, which can be used to plan mitigation actions against unknown attack scenarios. In addition, it is straightforward to define metrics for a subset of start nodes and a subset of end nodes similarly to Equation~\ref{eq:impact}.

\begin{figure*}[ht]
    \centering
        \begin{subfigure}{0.49\textwidth}
            \centering
        \includegraphics[width=1\textwidth]{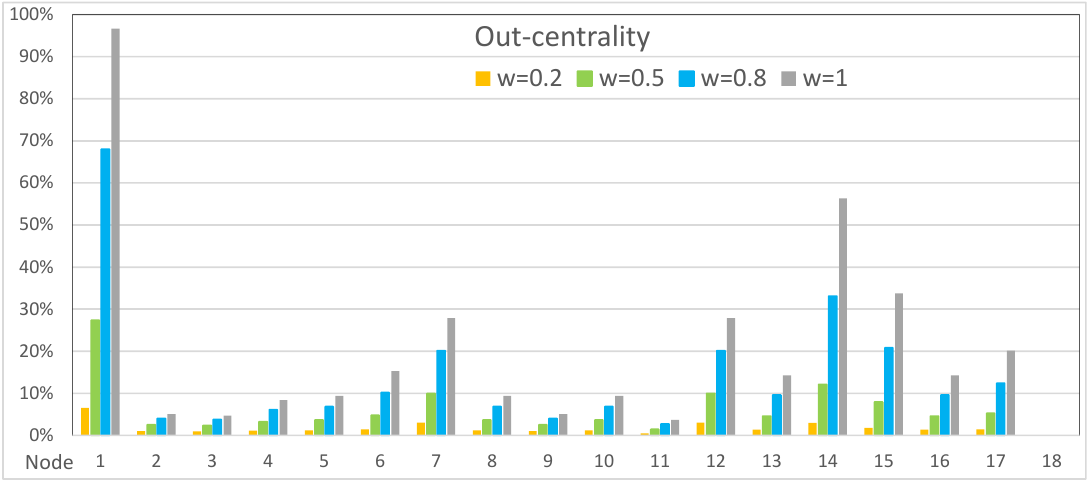}
        \label{fig:UC1Out}
    \end{subfigure}
        \begin{subfigure}{0.49\textwidth}
            \centering
        \includegraphics[width=1\textwidth]{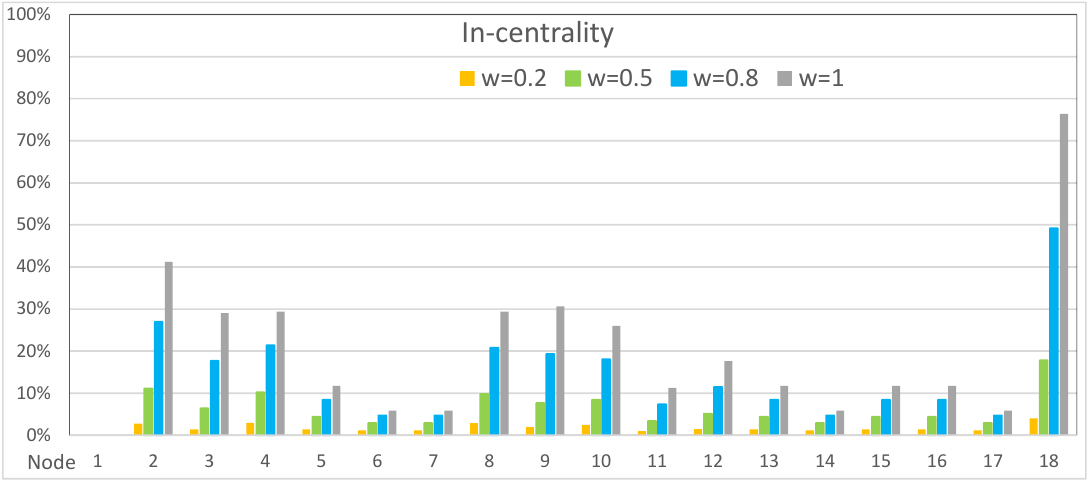}
        \label{fig:UC1In}
    \end{subfigure}
    \caption{Out- and in-centrality values of the Multi-cloud Enterprise Network of $18$ nodes in Figure~\ref{fig:MCE} from Eqs.~\ref{eq:out} and \ref{eq:in}, respectively. The histograms show the effects of using weighted link values according to the exploitability of vulnerabilities. Link weights are calculated by multiplying the Exploitability values from Table~\ref{tab:MCE}, which are based on the CVSS score values, with the scaling factors $w$.}
    \label{fig:UC1}
\end{figure*}

The out-centrality (Equation~\ref{eq:out}) and in-centrality (Equation~\ref{eq:in}) values for the $18$ nodes of the Multi-cloud network are shown in Figure~\ref{fig:UC1}. The results are shown for the four link weight values $w=0.2, 0.5, 0.8,$ and $1$. As the link weights are interpreted as probabilities of successfully exploiting vulnerabilities, lower link weight values correspond to reduced probabilities.

The out-centrality and in-centrality values depicted in Figure~\ref{fig:UC1} illustrate the typical characteristics of directed acyclic networks. Node $1$ (Start) has the highest out-centrality, while node $18$ (End) has the highest in-centrality. The out-centrality values tend to be more varied than the in-centrality values, as structures near the source nodes and high link weights boost propagation more than structures near target nodes. We also observe that the number of potential alternative paths affects the out-centrality and in-centrality values. Nodes $7, 12, 14,$ and $15$ have high out-centrality values because paths originating from these nodes have many alternatives. Similarly, multiple alternative paths leading to an end node increase its in-centrality value.

When planning, the mitigation actions are optimised based on detailed network modelling and the order of actions can change depending on the exploitability or link weight values of vulnerabilities. Additionally, skilled attackers may utilise longer attack path lengths because their probability of success remains high even with multiple consecutive exploits.

\paragraph{\textbf{Use Case 2: Netflix OSS architecture.}}

\begin{figure*}[ht]
    \centering
    \includegraphics[width=1.0\textwidth]{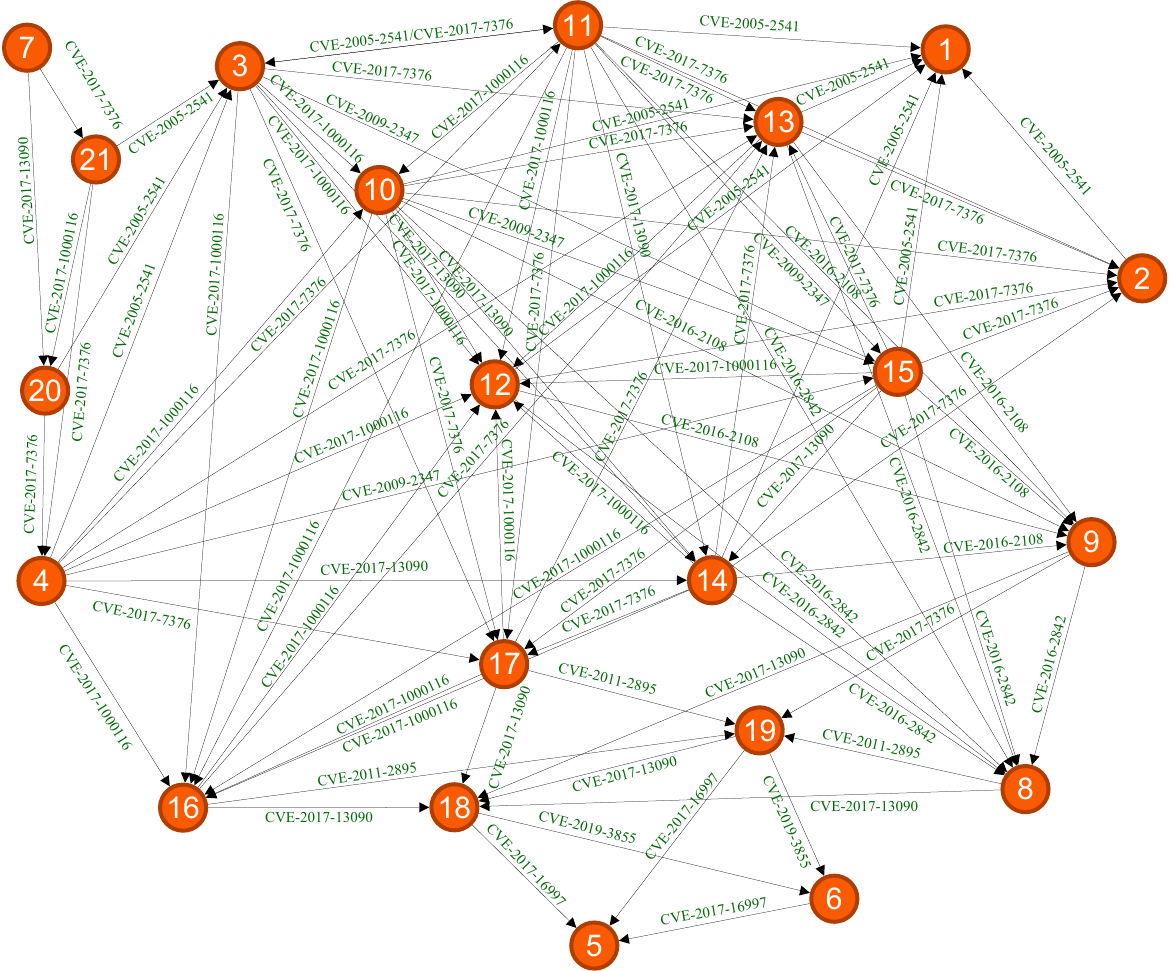}
    \caption{Use Case 2: Attack graph NetflixOSS from (\cite{stergiopoulos2022automatic}). Nodes are described in Table~\ref{tab:OSS_Nodes}.}
    \label{fig:OSS}
\end{figure*}

The second use case is a combination of containers provided by Netflix, consisting of the Spring cloud ecosystem. (\cite{stergiopoulos2022automatic}) Table~\ref{tab:OSS_Nodes} provides descriptions of the $21$ nodes in the NetflixOSS attack graph, the topology of which is depicted in Figure~\ref{fig:OSS}.

\begin{table*}[h]
    \centering
    \caption{List of nodes in the NetflixOSS attack graph in Figure~\ref{fig:OSS}.}
    \begin{tabular}{|c|c|l|l|} \hline 
 Node&Description & Node&Description \\ \hline \hline 
         1& Config service (Admin priv)
 & 12
&Service b (Admin priv)
\\ \hline 
         2& Config service (User priv)
 &  
13
& 
Service b (User priv)
\\ \hline 
         3& Eureka (Admin priv)
 &  
14
& Service c (Admin priv)
\\ \hline 
         4& Eureka (User priv)
 &  

15& 
 Service c (User priv)
\\ \hline 
         5& Hystrix dashboard (Admin priv)
 &  
16& Spring cloud dasboard (Admin priv)
\\ \hline 
         6& Hystrix dashboard (User priv)
 &  

17& 
 Spring cloud dasboard (User priv)
\\ \hline 
         7& Outside (Admin priv)
 &  
18& Turbine (Admin priv)
\\ \hline 
         8& rabbitmq (Admin priv)
 &  

19& 
 Turbine (User priv)
\\ \hline 
         9& rabbitmq (User priv)
 &  
20& Zuul (Admin priv)
\\ \hline 
 10&Service (Admin priv)
 &  

21& 
 Zuul (User priv)
\\\hline 
         11
& Service (User priv)
&  
& \\\hline
    \end{tabular}
    \label{tab:OSS_Nodes}
\end{table*}

\begin{figure*}[]
    \begin{subfigure}{0.5\textwidth}
        \centering
        \includegraphics[width=1\textwidth]{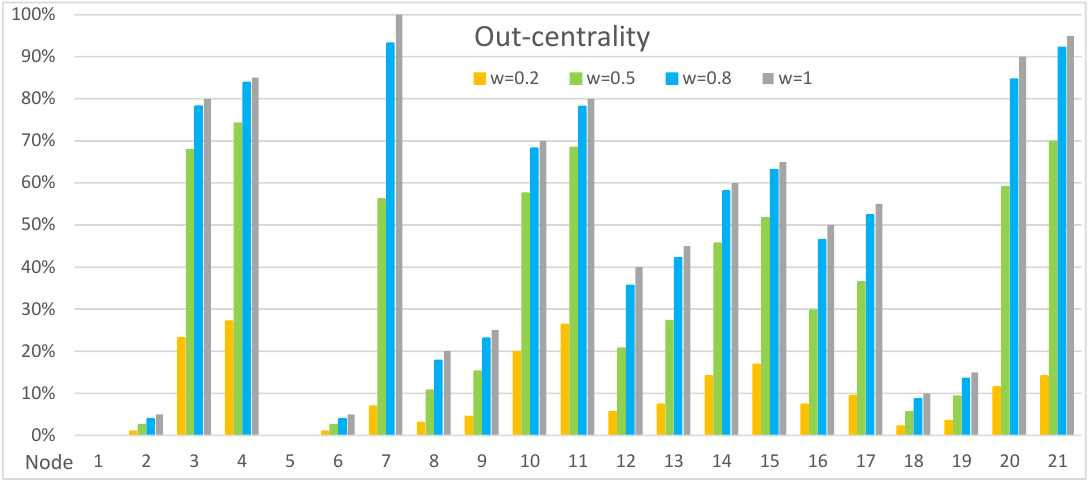}
        \label{fig:outOSS}
    \end{subfigure}
    \begin{subfigure}{0.5\textwidth}
        \centering
        \includegraphics[width=1\textwidth]{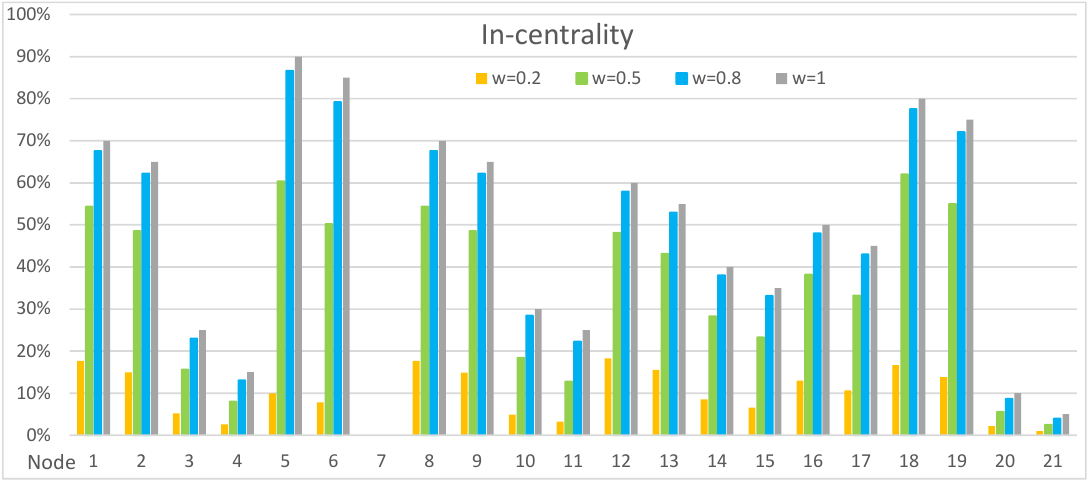}
        \label{fig:inOSS}
    \end{subfigure}
    \caption{Out-centrality and in-centrality values for the $21$ nodes of the NetflixOSS example network network in Figure~\ref{fig:OSS} for four different link weight values.}
     \label{fig:histOSS}
\end{figure*}

In Figure~\ref{fig:histOSS}, we show the out-centrality and in-centrality values of the NetflixOSS attack graph (\cite{stergiopoulos2022automatic}) for four link weight values, and they are found to increase monotonically as the link weight increases. However, the growth rate is slow for low link weights, then increases, and finally slows down again with high link weights due to saturation effects. CVSS scoring~(\cite{Scoring}) values are not used in this use case as they have almost equal values in this scenario.

The out-centrality and in-centrality in Figure~\ref{fig:histOSS} illustrate the two aspects of the NetflixOSS attack graph. The out-centrality values for the $21$ nodes of the graph indicate that nodes $3, 4, 7, 10, 11, 20,$ and $21$ are positioned at the beginning of possible attack chains, while nodes $1, 2, 5, 6, 8, 9, 18,$ and $19$ are at the end of the attack chains. The node $7$ is identified as the start node of the graph as it has a hundred per cent out-centrality value for the link weight values $w=1$. Nodes $12, 13, 14, 15, 16,$ and $17$ are in the intermediary positions within the attack chains, which can be used to prioritise protective actions or mitigate the impacts of the ongoing cyber attacks.

\begin{figure}[ht]
    \centering
        \includegraphics[width=0.5\textwidth]{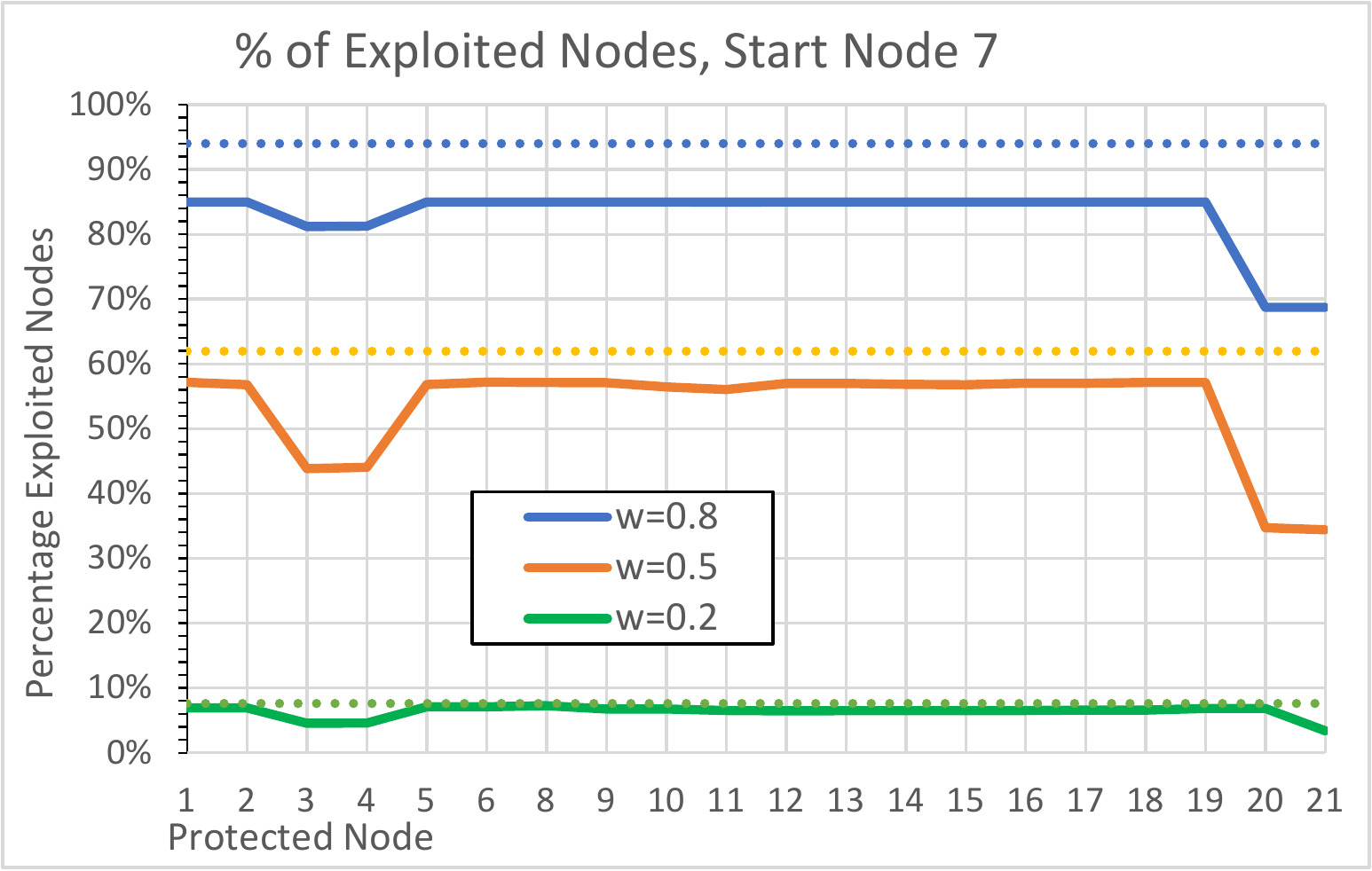}
        \caption{The average percentage values of exploited nodes in the NetflixOSS example network in Figure~\ref{fig:OSS} when a node is protected. The effects are calculated as average effects on other nodes. The results are shown for link weight values of $w=0.2$, $w=0.5$, and $w=0.8$. The start node of the attack is $7$. The dotted lines show the average percentage of exploited nodes when nodes are not protected.}
    \label{fig:OSS21a}
\end{figure}

In Figure~\ref{fig:OSS21a}, the average percentage of exploited nodes is shown when a node's vulnerabilities are mitigated. The results are displayed for link weight values $w=0.2$, $w=0.5$, and $w=0.8$ when the start node is $7$. Here the effects are computed as average effects on other nodes. The dotted lines illustrate the average percentage of exploited nodes when the nodes are not protected. The relationship between the link weights of the nodes and the average effect on the network is not linear. For instance, the link weight of $w=0.2$ results in only about 10 per cent exploitation in the network. Meanwhile, the link weights of $w=0.5$ and $w=0.8$ lead to approximately sixty and ninety per cent exploitation levels, respectively.

A significant decrease in the number of exploited nodes is seen when the vulnerabilities of the 
nodes $3, 4, 20,$ and $21$ in the graph are mitigated. This is expected because these nodes are at the beginning of potential attack chains starting from node $7$. The impact is bigger when the links have higher weights. Nodes that come right after the start node become more important when the link weight values (exploitability) increase.

\balance
When it comes to addressing vulnerabilities, it may be more useful to focus on eliminating vulnerabilities related to services rather than just protecting individual nodes in the attack graph. Next, we compare how out-centrality and in-centrality values decrease when vulnerabilities in services are protected. The comparison will be performed by taking the difference between the unprotected and protected results. Next, we calculate out-centrality (Equation~\ref{eq:out}) and in-centrality (Equation~\ref{eq:in}) results for all nodes in the attack graph in Figure~\ref{fig:OSS}. In these calculations, we assume that the cyber attack starts at node $s$ in Equation~\ref{eq:out} or nodes $t$ in Equation~\ref{eq:in} with certainty. If the attack starts from the first node in the attack graph in Figure~\ref{fig:OSS}, then the relevant results focus solely on node $7$.

\begin{figure*}[ht]
    \centering
    \begin{subfigure}{1\textwidth}
        \includegraphics[width=\textwidth]{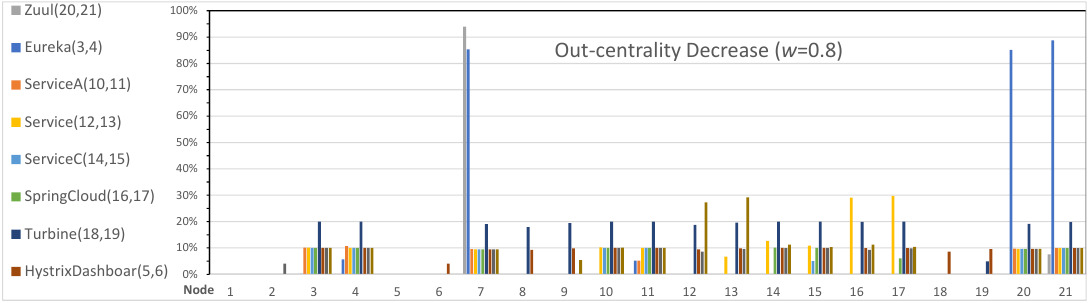}
        \label{fig:OSSOutEro}
    \end{subfigure}
    \begin{subfigure}{1\textwidth}
        \includegraphics[width=\textwidth]{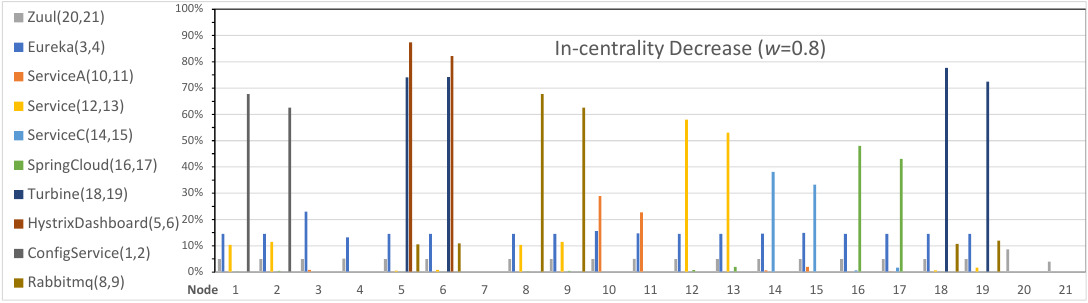}
        \label{fig:OSSInEro}
    \end{subfigure}
    \caption{Out-centrality and in-centrality differences (i.e. decrease) for nodes $1$ to $21$ in Figure~\ref{fig:OSS} after mitigating vulnerabilities in services indicated in the legend (link weight $w=0.8$).}
    \label{fig:OSScve}
\end{figure*}

Figure~\ref{fig:OSScve} shows these differences (i.e. decrease) for nodes $1-21$ after mitigating all the vulnerabilities in the services indicated in the legend. For this example, the link weight is $w=0.8$ to represent a skilled perpetrator. Looking at the out-centrality histogram in Figure~\ref{fig:OSScve}, we observe that four bars stand out: Zuul and Eureka services have decreased by 95\% and 85\% for the out-centrality of node $7$, and Eureka services also show a significant decrease for the out-centrality values of node $20$ (by 85\%) and for node $21$ (by 88\%).

The in-centrality values shown in Figure~\ref{fig:OSScve} indicate that protecting services at the nodes leads to a decrease in the in-centrality values of these particular nodes. For example, Turbine services have a significant impact on nodes $18$ and $19$. Furthermore, protecting Turbine services results in a substantial decrease in the in-centrality values of nodes $5$ and $6$. These effects result from nodes 18 and 19 denoting Turbine services states (see Table~\ref{tab:OSS_Nodes}), while nodes 5 and 6 represent subsequent Hystrix dashboard states in the attack graph (Figure~\ref{fig:OSS}). On the other hand,
protecting services at the beginning of attack chains, such as Zuul and Eureka services, affects the in-centrality values of nodes $3$ and $4$. Services exploited later in the attack chains have no effect on these nodes.

\begin{figure*}[h]
    \centering
        \begin{subfigure}{1\textwidth}
        \includegraphics[width=1\textwidth]{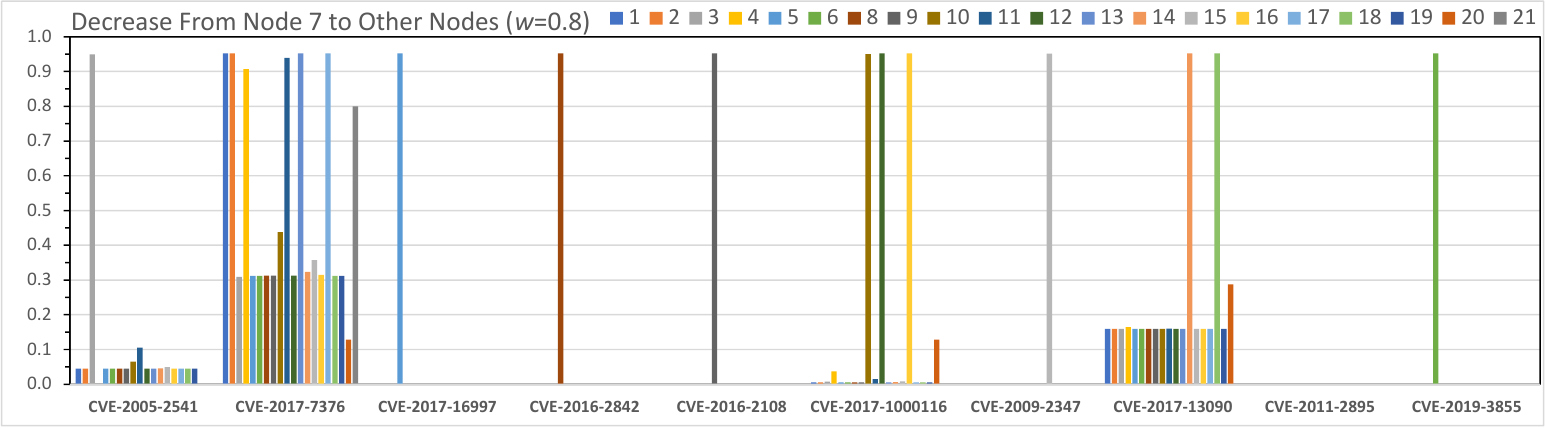}
        \caption{Probability differences (decrease) of the in-centralities from the attack start node $7$ to other nodes (link weight $w=0.8$) after mitigating a vulnerability from the network. Vulnerabilities are indicated on the horizontal axis.}
    \label{fig:OSScve7}
    \end{subfigure}
    \begin{subfigure}{1\textwidth}
        \includegraphics[width=1\textwidth]{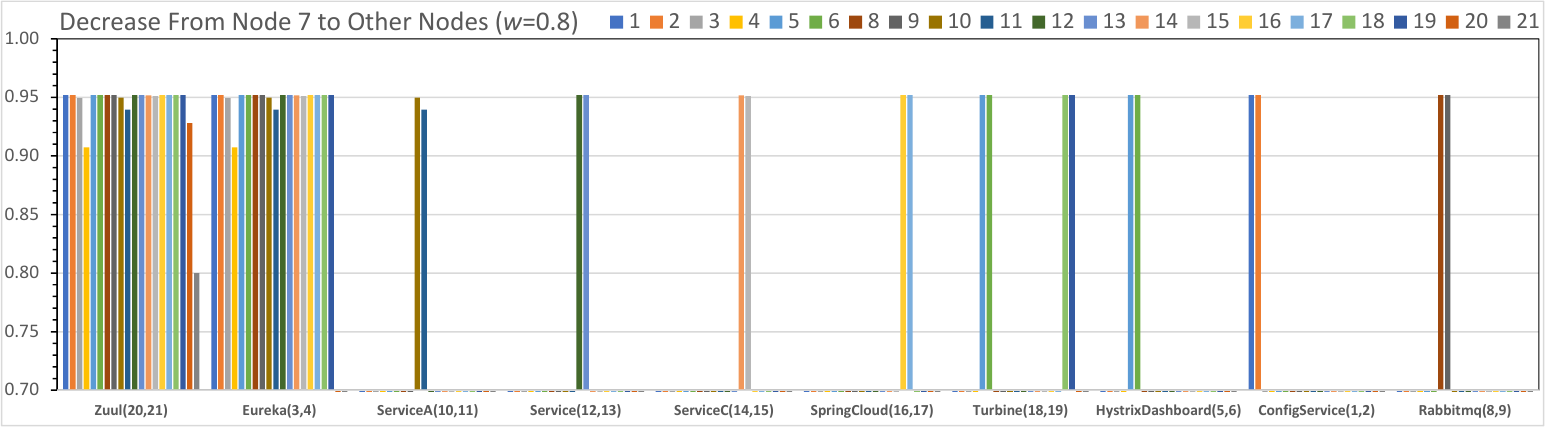}
        \caption{Probability differences (decrease) of the in-centralities from the attack start node $7$ to other nodes (link weight $w=0.8$) after removing all vulnerabilities from a server. Servers are indicated on the horizontal axis.}
    \label{fig:OSSPalv7}
    \end{subfigure}
        \caption{Summaries generated from the model (a) when the entire network is protected against a single vulnerability or (b) a server is protected against all its vulnerabilities.}
        \label{OSScvepalv}
\end{figure*}

Figure~\ref{fig:OSScve7} illustrates the decrease of the in-centrality values from the start node $7$ to other nodes in the graph for the link weight of $w=0.8$ after mitigating a vulnerability indicated on the horizontal axis. In the model (see Section~\ref{sec:Model}) these correspond to a decrease in the matrix element values $C(s,t)$, for $s=7$ and $t=1, 2, 3, 4, 5, 6, 8, ..., N$. By mitigating the vulnerabilities CVE-2017-7376 and CVE-2017-13090, a significant decrease in the in-centrality values for several nodes can be achieved. Results at the node level indicate that CVE-2017-7376 significantly affects the in-centralities of nodes $1, 2, 4, 11, 13, 17,$ and $21$, with smaller effects on the other nodes. On the other hand, the vulnerability CVE-2017-13090 has a major impact on nodes $14$ and $18$, but with minor effects on the other nodes.

\begin{figure*}[h]
    \centering
        \begin{subfigure}[t]{0.4\textwidth}
        \includegraphics[width=\textwidth]{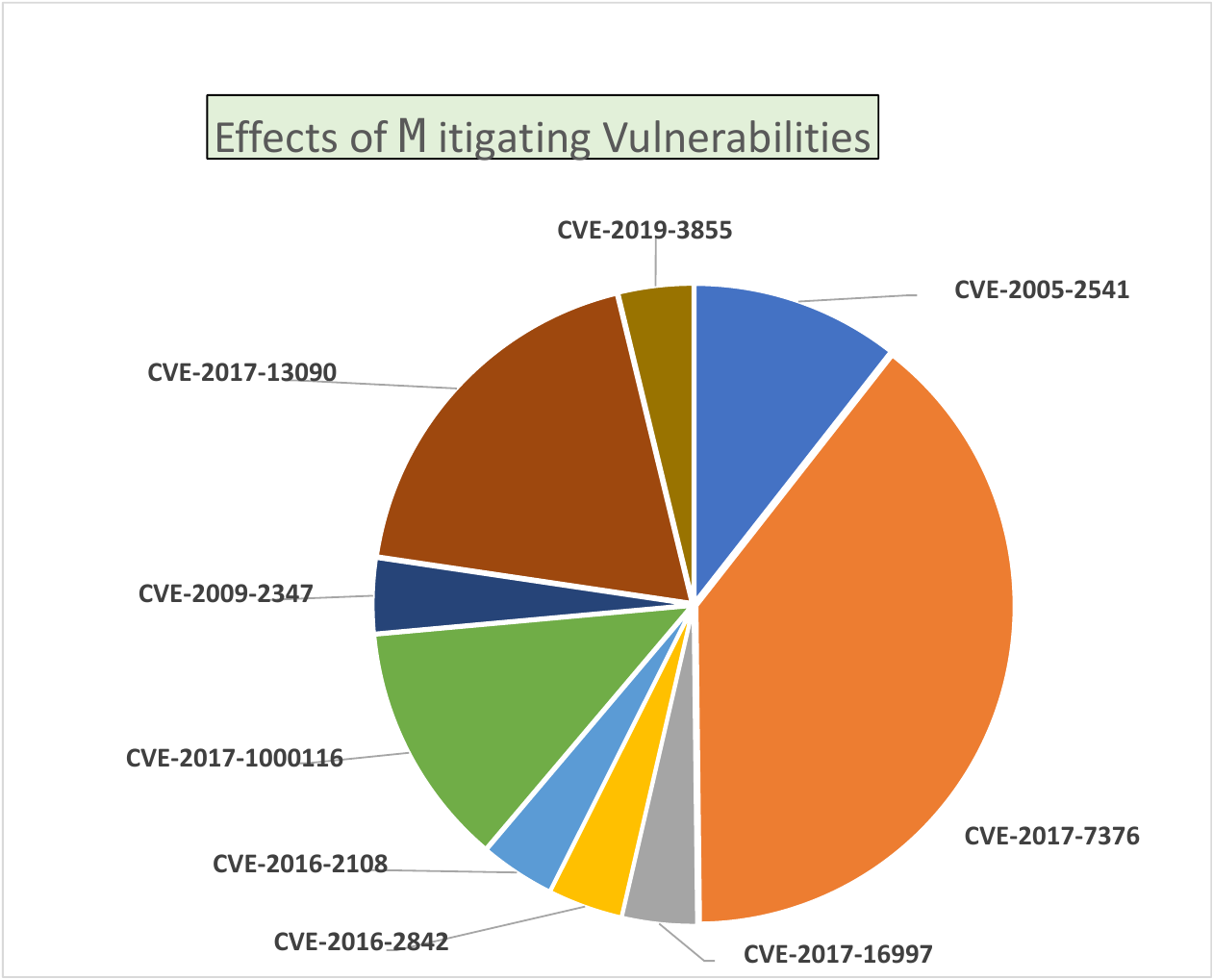}
        \caption{A specific vulnerability removed from the attack graph indicated on the legend (Summary from Figure~\ref{fig:OSScve7}).}
    \label{fig:TotOut7}
    \end{subfigure}
        \begin{subfigure}[t]{0.4\textwidth}
        \includegraphics[width=\textwidth]{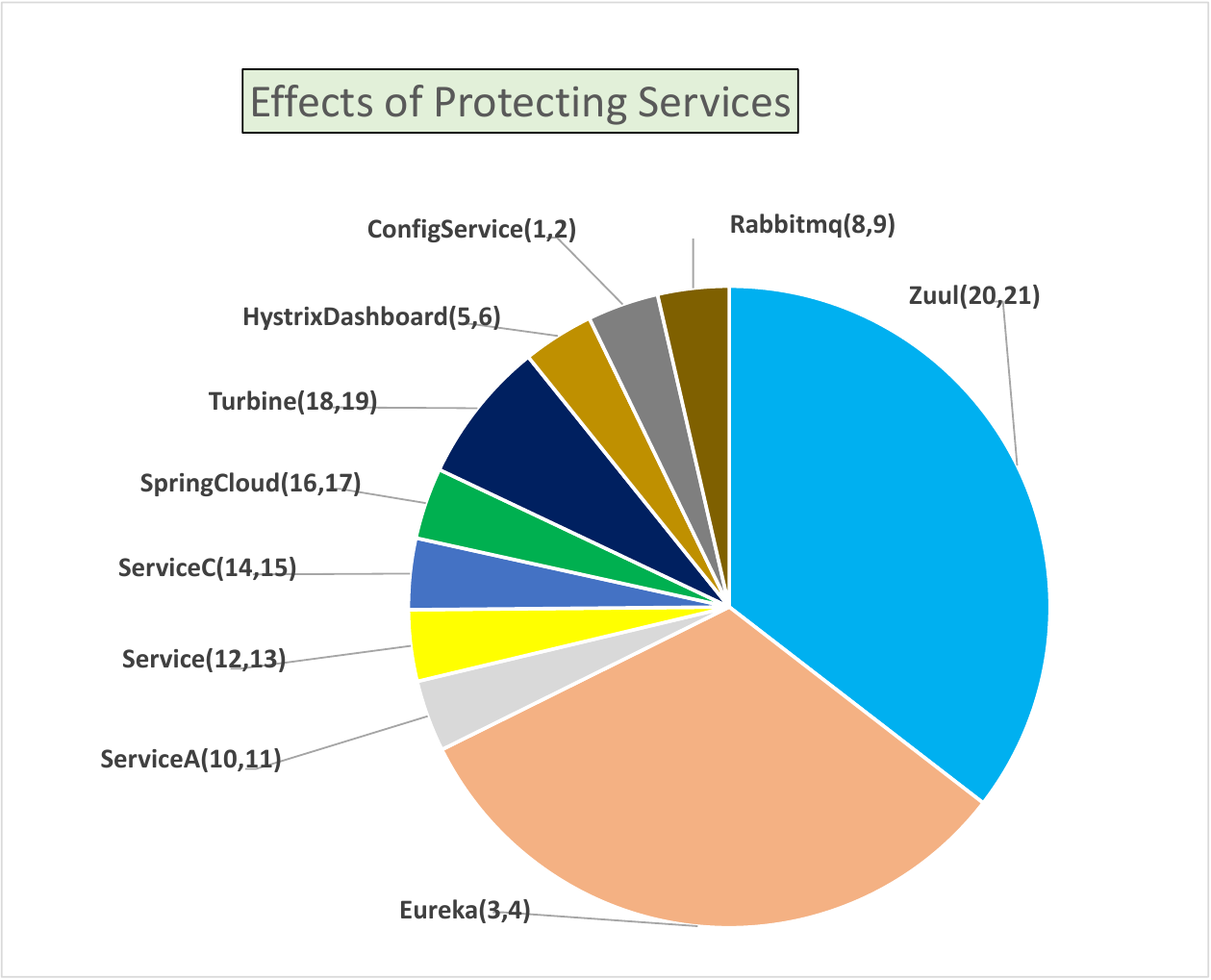}
        \caption{Vulnerabilities removed from specific servers/services indicated on legend (Summary from Figure~\ref{fig:OSSPalv7}).}
    \label{fig:TotPalvOut7}
    \end{subfigure}
        \caption{Relative effects after removing a vulnerability from the network or all vulnerabilities from a server. As an example, the results are shown for the link weight value $w=0.8$.}
        \label{TotPalv7}
\end{figure*}

Figure~\ref{fig:OSSPalv7} illustrates the decrease in the in-centrality values from the start node $7$ to other nodes in the graph for the link weight of $w=0.8$ after mitigating vulnerabilities in the services indicated on the horizontal axis. Zuul and Eureka are the most effective services to be protected, as they are at the beginning of attack chains.
In Figure~\ref{fig:TotOut7} we show 
the summary of the decrease in the average out-centrality from the start node $7$ to other nodes for the link weight $w=0.8$ after mitigating the vulnerabilities. The four most important vulnerabilities in this order are CVE-2017-7376, CVE-2017-13090, CVE-2017-1000116, and CVE-2005-2541. Figure~\ref{fig:TotPalvOut7} shows the corresponding results for the services. The most important services are Zuul and Eureka as we have already seen in Figure~\ref{fig:OSSPalv7}.

\paragraph{\textbf{Use Case 3: Pony APT campaign.}}

The third use case involves a causal graph generated in (\cite{alsaheel2021atlas}) using an attack investigation tool. This tool integrates natural language processing and deep learning techniques into data analysis to model sequence-based attack and non-attack entities.

\begin{figure*}[h]
    \centering
    \begin{subfigure}[b]{0.75\textwidth}
    \centering 
    \includegraphics[width=1\textwidth]{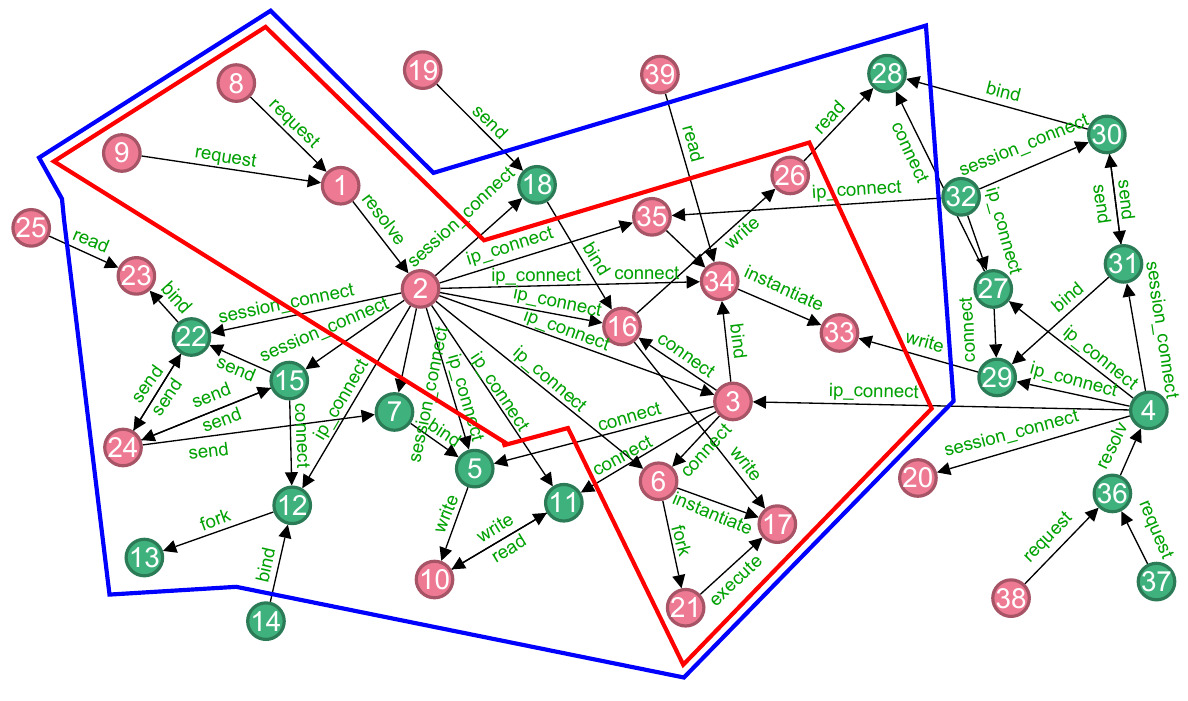}
    \end{subfigure}
    \begin{subfigure}[b]{1\textwidth}
    \centering
    \includegraphics[width=1\textwidth]{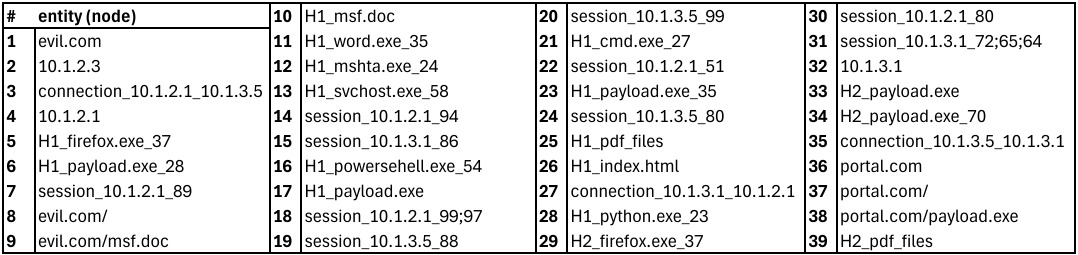}
    \end{subfigure}
    \caption{Use Case 3: Attack entities are identified based on the connections within the graph structure. The original attack entities are highlighted in red, while the original non-attacks are marked in green. The blue line represents the extended set of attack nodes detected by our network modelling method, whereas the red line indicates the minimum set of reachable nodes confined to the original set of attack nodes. The list of nodes is provided in the table from (\cite{alsaheel2021atlas}).}
    \label{fig:AtlasYhteys}
\end{figure*}

Figure~\ref{fig:AtlasYhteys} depicts attack and non-attack entities detected by the ATLAS investigation tool. (\cite{alsaheel2021atlas}) The original attack entities are indicated by the red colour and the original non-attack entities by the green colour. (\cite{alsaheel2021atlas}) The blue line shows the extended set of attack nodes reachable from start nodes $8$ or $9$ by following all alternative paths in the graph. The red line shows the minimum set of nodes restricted to the original set of attack nodes.

\begin{figure*}[ht]
    \centering
    \begin{subfigure}{1\textwidth}
        \centering
        \includegraphics[width=1\textwidth]{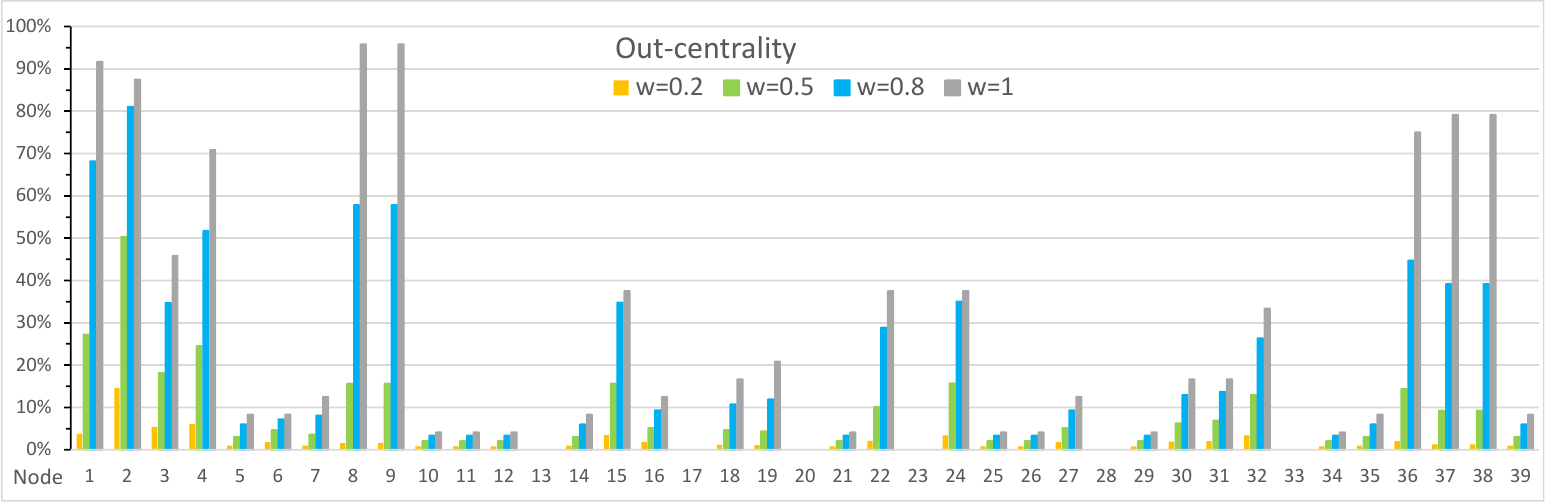}
    \end{subfigure}
    \begin{subfigure}{1\textwidth}
        \centering
        \includegraphics[width=1\textwidth]{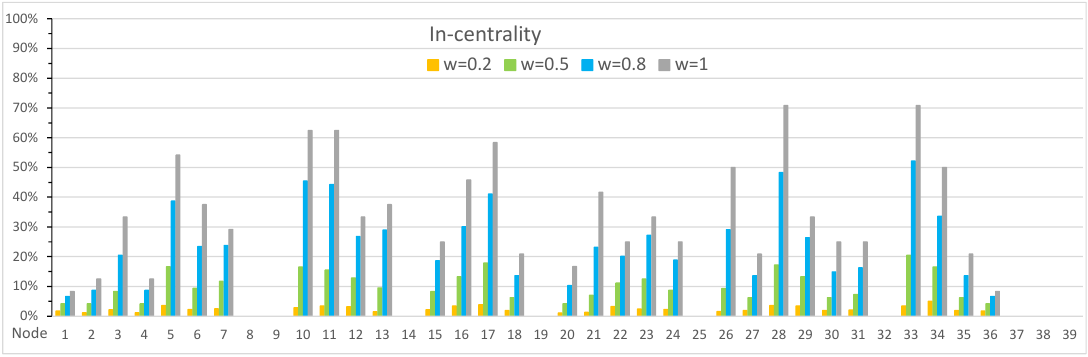}
    \end{subfigure}
    \caption{Out-centrality and in-centrality values for the 39 nodes of the ATLAS causal graph in Figure~\ref{fig:AtlasYhteys} for four different link weight values.}
    \label{fig:histAtlas}
\end{figure*}

In Figure~\ref{fig:histAtlas} we show the out- and in-centrality values for the $39$ nodes of the ATLAS causal graph for four different link weight values. The histograms in Figure~\ref{fig:histAtlas} are similar to those in Figure~\ref{fig:histOSS}, but there are some differences. When comparing the ATLAS and NetflixOSS graphs, we can observe that the out-centrality and in-centrality values for the ATLAS causal graph for low link weight values are much smaller. Additionally, the centrality values for higher link weight values are roughly half of the Netflix results. These differences are attributed to the higher density and higher mean node degree of the NetflixOSS graph compared to the ATLAS graph.

\begin{figure*}[h]
    \begin{subfigure}{1\textwidth}
        \includegraphics[width=0.49\textwidth]{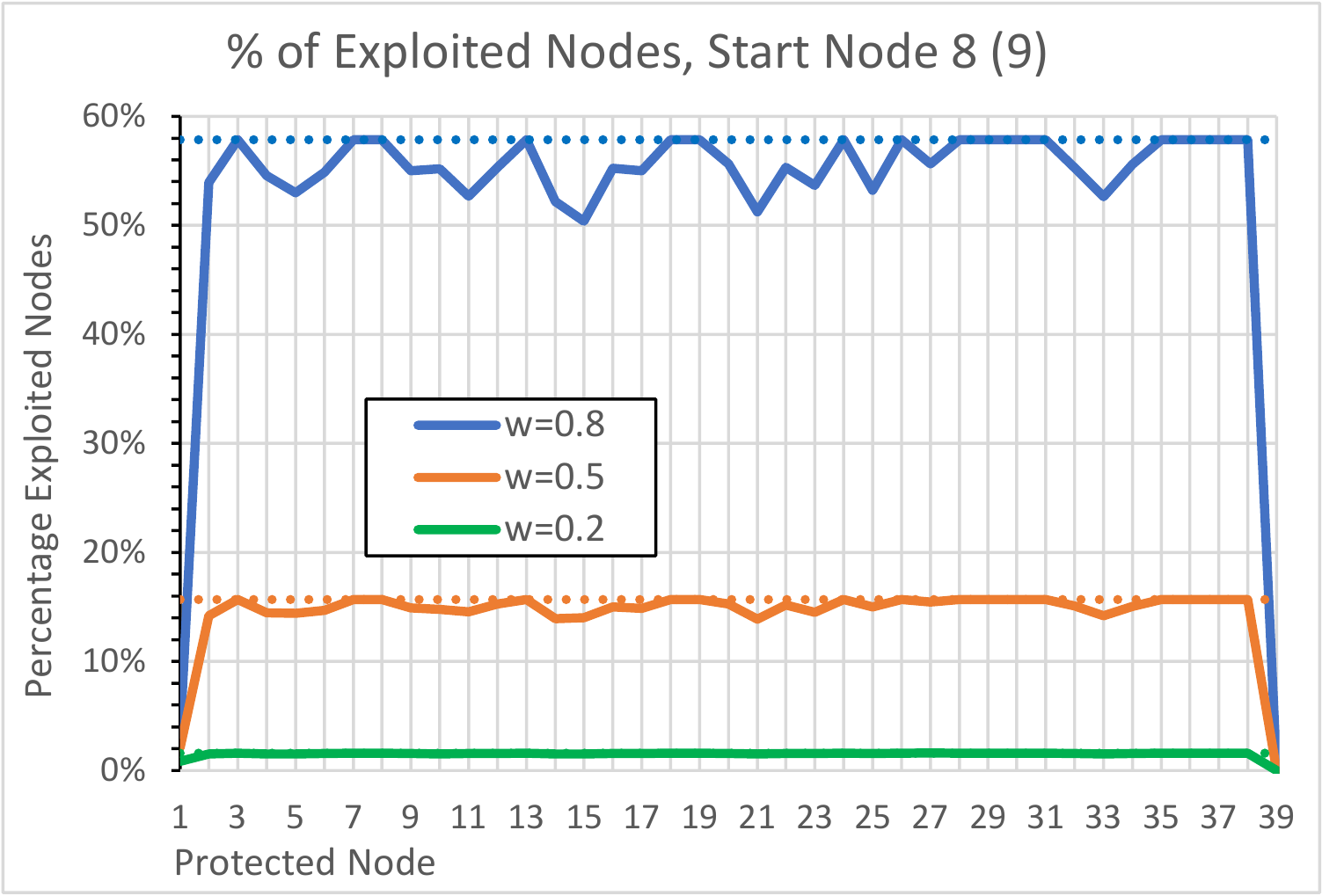}
        \centering
        \end{subfigure}
        \begin{subfigure}{1\textwidth}
        \includegraphics[width=0.49\textwidth]{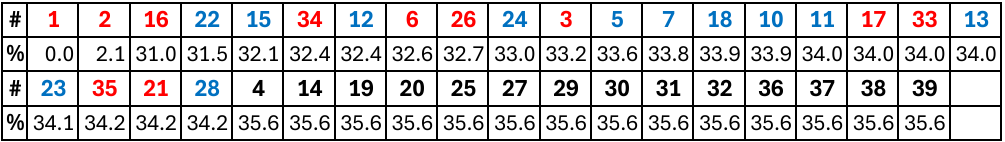}
        \centering
        \end{subfigure}
        \caption{The average percentage values of exploited nodes in the ATLAS causal graph in Figure~\ref{fig:AtlasYhteys} when a node is protected. The effects are calculated as average effects on other nodes. The results are shown for the link weight values $w=0.2$, $w=0.5$, and $w=0.8$. The start node is $8$ or $9$. The dotted lines show the average percentage values of exploited nodes when nodes are not protected. In the table, nodes are ordered according to their criticality for $w=0.8$ (blue curve).}
    \label{fig:Atlas39a}
\end{figure*}

Figure~\ref{fig:Atlas39a} illustrates the percentage of exploited nodes in the ATLAS causal graph when one of the $39$ nodes is protected. The results are presented for three link weight values: $w=0.2$, $w=0.5$, and $w=0.8$. The start node is node $8$ or node $9$. The effects are calculated as average values on the other nodes. The dotted lines represent the average percentage of the exploited nodes when no nodes are protected. Compared to Figure~\ref{fig:OSS21a} results, we observe more fluctuations but with smaller
amplitude. Mitigating vulnerabilities in the nodes $1$, $2$, $16$, $15$, $34$ or $12$ has the largest effects of protecting against the attack. Similarly to the case of Figure~\ref{fig:histAtlas}, the lower density causes the curves to be lower compared to those of Figure~\ref{fig:OSS21a}, and the ratios between the curves turn out to be different.

Use case 3 illustrates how our model can analyse incomplete graph information. The original attack graph produced by the investigation tool (\cite{alsaheel2021atlas}) is indicated by the red line in Figure~\ref{fig:AtlasYhteys}. However, more nodes can be accessed from starting 
nodes $8$ or $9$. Consequently, the original attack graph can be extended by including these additional nodes, some of which have been classified as non-attack nodes (in green) by the investigation tool. This enlarged attack graph is shown by the blue line in Figure ~\ref{fig:AtlasYhteys}. Certain nodes remain outside of this node set, with five identified as attack nodes (in red). This suggests that the actual attack graph could even be larger, and it is important to also consider these outlying nodes of Figure~\ref{fig:AtlasYhteys}.

Figure~\ref{fig:AtlasYhteys} also presents a table that lists the nodes together with their effects when protected. As an example, the link weight value of $w = 0.8$ is used for all links in the graph. When an attack starts from node $8$ or $9$, the average percentage of exploited nodes in the graph is $35.6\%$ when no nodes are protected. When node $1$ is protected, $0.0\%$ of the nodes are exploited, and when node $2$ is protected, $2.1\%$ are exploited. This outcome is attributed to the critical positions of these nodes within the graph structure. In the table of Figure~\ref{fig:AtlasYhteys}, the tool's classified attack nodes are indicated in red, non-attack nodes in blue, and non-reachable nodes (both attack and non-attack) in black. When prioritising mitigation measures, nodes $22$ and $15$ have significant effects on 
the extended graph, although they have not been recognised as attack nodes by the investigative tool.

\begin{figure*}[ht]
    \centering
    \begin{subfigure}{1\textwidth}
        \includegraphics[width=\textwidth]{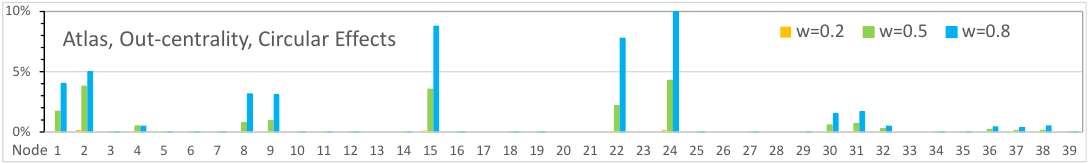}
    \end{subfigure}
    \begin{subfigure}{1\textwidth}
        \includegraphics[width=\textwidth]{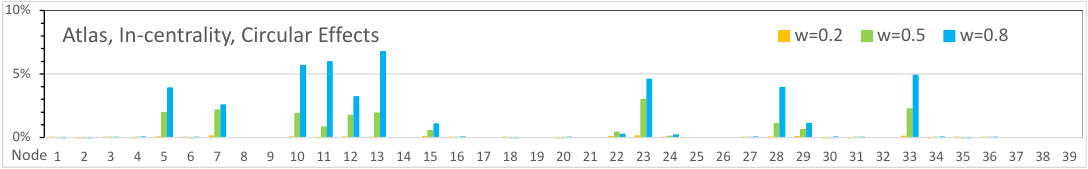}
    \end{subfigure}
    \caption{Circular Effects in the ATLAS causal graph of Figure\ref{fig:AtlasYhteys} for three different link weight values. The bars in the histograms show differences between two calculations: full breakthrough propagation and self-avoiding paths between nodes.}
    \label{fig:CircularAtlas}
\end{figure*}

In contrast to the networks in use cases 1 and 2, the ATLAS causal graph (Figure~\ref{fig:AtlasYhteys}) has a loop between the nodes $15, 22,$ and $24$, and bidirectional links between the nodes $10$ and $11$, and between the nodes $30$ and $31$. In our model, we assume that due to a loop and bi-directionality, circular and recurrent effects are possible. Figure~\ref{fig:CircularAtlas} illustrates the circular effects in the ATLAS causal graph for three different link weight values. The histogram shows a difference between two situations, namely one with circular effects and the other with only self-avoiding paths in the network structure. The effects vary according to the weight values of the links.
They are low when the link weights are $w=0.2$, peak at higher link weights, and gradually slow down due to saturation effects for high link weight values.

To summarise, it turns out that for the attack graph with one loop and two bidirectional links, the out-centrality values are increased. Furthermore, nodes $1, 2, 8,$ and $9$ have increasing out-centrality values because they are at the beginning of attack paths that can reach the loop nodes $15$, $22$ or $24$. Upon examining the graph in Figure~\ref{fig:Atlas39a}, we conclude that the in-centrality values of nodes $5$, $7$, $10$, $11$, $12$, $13$, and $23$ have increased due to the circular effects of the loop, and the in-centrality values of nodes $28$, $29$ and $33$ have increased due to the recurrent effects between nodes $30$ and $31$ by the two bidirectional links.

\section{Concluding Remarks}
\label{sec:Conclusions}
\let\newpage\relax
In cybersecurity analysis of computer networks and service systems, it is natural to take a graph-based approach. Until recently, cyber-related graphs, such as attack graphs, have been created manually or generated by scanning network configurations to identify vulnerabilities and compute potential attack paths. In this case, the analysis of an attack graph involves examining individual paths that an attacker could take and evaluating the risk associated with each path.

In the present study, we have introduced a novel probabilistic graph-based analysis approach to investigate the structural and dynamic properties of the attack and causal graphs. This approach has previously been used in our research to model the spread of influence in complex networks. Here, we are applying this methodology for the first time to analyse the propagation of attacks and causal influences. In this context, out-centrality and in-centrality measures are used to identify critical nodes in the graph. The documented exploitability values of vulnerabilities mimic the probability of propagation of attacks or causal influences on possible paths or sequences of actions that an attacker or influencer can follow to achieve a particular objective, such as compromising a critical asset in a network. Similarly, the documented impact values of vulnerabilities are used to demonstrate the model in analysing cumulative and individual impacts of attacks.

Future research could focus on multilayer modelling of enterprise infrastructure and attack graphs. Additional areas of interest could include the integration of real-time attack graph generation and the exploration of hybrid approaches that combine machine learning or anomaly detection methods.
Recently, there has emerged a variety of studies on AI-driven techniques in cybersecurity (\cite{salem2024advancing}) and intrusion detection (\cite{Yee}). These approaches offer new opportunities to be integrated with the probabilistic methods discussed in this study.

In summary, we have developed a new
method for analysing alternative paths within an attack graph and calculating their combined exploitability and impact. This approach differs from the previous methods since our approach offers a probabilistic framework to assess the combined effects of the entire attack graph, rather than focusing solely on the maximum or minimum scenarios.  We have applied our model by presenting results related to functional services based on identified vulnerabilities. Additionally, this method can be extended to analyse multiple attack graphs simultaneously in a network structure. For higher levels of abstraction and summaries, it is essential to consistently aggregate alternative attack or causal effect paths within the model. This integrated perspective improves understanding of the situation and helps prioritise mitigation efforts. Thus, our model could be a versatile tool in the hands of cyber analysts.

\section*{Data Availability}\label{sec:DataAvailability}
The datasets utilised for the use cases of this article had undergone several preprocessing steps and have been obtained from open online sources. The respective datasets and graphs have been openly provided by the authors of the referenced original articles (\cite{stergiopoulos2022automatic, ibrahimAttackGraphGeneration2019, alsaheel2021atlas}). 

\printbibliography

\end{document}